\documentclass[letterpaper, 10 pt, conference]{ieeeconf}  

\IEEEoverridecommandlockouts                              

\usepackage{graphicx}      
\usepackage{algorithm}     
\usepackage{amsmath} 
\usepackage{amsfonts}
\usepackage{tabularx}
\usepackage{amssymb} 

\usepackage{algpseudocode}
\usepackage{xcolor}

\usepackage{longtable}
\usepackage{makecell}

\usepackage{comment} 

\title{\LARGE \bf
Sentiment-driven prediction of financial returns:\\ a Bayesian-enhanced FinBERT approach
}

\author{Raffaele Giuseppe Cestari and Simone Formentin
\thanks{This paper is partially supported by FAIR (Future Artificial Intelligence Research) project, funded by the NextGenerationEU program within the PNRR-PE-AI scheme (M4C2, Investment 1.3, Line on Artificial Intelligence), by the Italian Ministry of Enterprises and Made in Italy in the framework of the project 4DDS (4D Drone Swarms) under grant no. F/310097/01-04/X56 and by the PRIN PNRR project  P2022NB77E “A data-driven cooperative framework for the management of distributed energy and water resources” (CUP: D53D23016100001), funded by the NextGeneration EU program.}
\thanks{The authors are with Department of Electronics, Information, and Bioengineering, Politecnico di Milano,
Milano, Italy. Email to: {\tt\small raffaelegiuseppe.cestari@polimi.it}.}%
}

\begin{document}

\maketitle
\thispagestyle{empty}
\pagestyle{empty}

\begin{abstract}

Predicting financial returns accurately poses a significant challenge due to the inherent uncertainty in financial time series data. Enhancing prediction models' performance hinges on effectively capturing both social and financial sentiment. In this study, we showcase the efficacy of leveraging sentiment information extracted from tweets using the FinBERT large language model. By meticulously curating an optimal feature set through correlation analysis and employing Bayesian-optimized Recursive Feature Elimination for automatic feature selection, we surpass existing methodologies, achieving an F1-score exceeding 70\% on the test set. This success translates into demonstrably higher cumulative profits during backtested trading. Our investigation focuses on real-world SPY ETF data alongside corresponding tweets sourced from the StockTwits platform.

\end{abstract}

\section{INTRODUCTION}

Predicting the sign of financial asset returns with precision remains a challenging task (\cite{cestari2023hawkes}). Given the human-centric nature of finance, textual data related to financial topics represents valuable information that can enhance prediction models. In the literature, Natural Language Processing (NLP) techniques have been extensively employed for sentiment extraction, employing deep learning models \cite{xu2019stock,liu2023stock} as well as lexicon-based approaches \cite{koukaras2022stock}. Notably, the VADER (Valence Aware Dictionary and sEntiment Reasoner) algorithm \cite{hutto2014vader} has been utilized for sentiment classification, demonstrating its efficacy in predicting price movements, as exemplified in the case of Microsoft stock.

A comprehensive review of sentiment analysis techniques in the financial domain is provided by \cite{8780312}. Gupta and coauthors in \cite{gupta2020sentiment} emphasize the importance of sentiment data volume in enhancing prediction accuracy, highlighting financial tweets sourced from platforms like Twitter (now X) or StockTwits as primary data sources \cite{mittal2012stock,pagolu2016sentiment}.

In this study, \textit{we investigate the influence of both model selection and input features on prediction quality} using real StockTwits tweets associated with SPY ETF financial time series. We begin with a feature set outlined in \cite{liu2023stock} and employ an automatic feature selection algorithm, Bayesian-optimized Recursive Feature Elimination (BO-RFE), to extract the most informative subset of features. This process combines Recursive Feature Elimination (RFE) with Bayesian Optimization (BO) to iteratively converge on the optimal feature set, evaluated based on the F1-score.
Additionally, \textit{we conduct a correlation analysis to elucidate the dynamic impact of sentiment}, as classified by FinBERT \cite{araci2019finbert}, \textit{on returns}. This analysis enriches the set of regressors and aids in top-feature selection.

We compare three architectures, employing the same prediction model but varying in input features and training window size:
\begin{itemize}
\item \textit{Literature}: Benchmark architecture as described in \cite{liu2023stock}.
\item \textit{BO-RFE-2} (2 regressors): Literature architecture with a reduced set of input features obtained from BO-RFE.
\item \textit{BO-RFE-5} (5 regressors): Literature architecture with a reduced set of input features obtained from BO-RFE, augmented with additional correlation-based regressors.
\end{itemize}

Our findings demonstrate that:
\begin{itemize}
\item Utilizing BO-RFE/correlation-based feature selection with the same prediction model (Support Vector Machine, SVM) yields superior results compared to the benchmark \cite{liu2023stock} in terms of F1-score (above 70\%) and simulated profits in real trading days.
\item Correlation analysis significantly contributes to feature selection by capturing slow sentiment dynamics.
\item Sentiment-based features play a crucial role, as evidenced by the shift from 3 out of 10 sentiment features in the benchmark to 4 out of 5 in our proposed BO-RFE-5 architecture.
\item Reducing features to essential elements enables a shorter training window, resulting in better results with less data and ensuring model reactivity.
\end{itemize}

The remainder of this paper is organized as follows: Section \ref{data} presents the characteristics of the dataset, followed by Section \ref{sentiment} detailing the FinBERT classification of tweets and sentiment index computation. Section \ref{BORFESECT} outlines the automatic feature selection procedure, while Section \ref{correlation} discusses the correlation analysis and the enriched set of candidate optimal regressors. Section \ref{prediction} describes the prediction model structure, and Section \ref{numericalSim} presents the numerical simulations and results, including return sign prediction metrics on the test set (Section \ref{accuracy}) and simulated profits in actual trading days (Section \ref{simulation}).

\section{THE DATASET}
\label{data}
The data on which this article is based are offered by \cite{liu2023stock}. 
The sources of information are:
\begin{itemize} 
\item The financial time series of the SPDR S\&P 500 ETF (SPY), a fund replicating the S\&P 500 index including the IT, finance, energy, and TLC sectors (the 500 largest US companies). It implements automatic share balancing based on stock value.
\item A collection of $3,261,867$ tweets gathered through StockTwits API, tagged SPY, containing the users' sentiment information on the same SPY asset. 
\end{itemize}
The temporal horizon is from 2020-03-23 to 2022-03-14. To avoid redundancy, please refer to \cite{liu2023stock} for further details.

\section{TWEETS CLASSIFICATION AND SENTIMENT SCORING}
\label{sentiment}

FinBERT, introduced by Araci et al. \cite{araci2019finbert}, is a specialized variant of the Google BERT (Bidirectional Encoder Representations from Transformers) model \cite{devlin2018bert}, specifically tailored for financial text analysis. BERT itself is a sophisticated language model built upon the transformer architecture, featuring 12 encoders with 12 bidirectional self-attention heads, constituting either 110 million parameters or, in its larger variant, 24 encoders with 16 bidirectional self-attention heads, amounting to 340 million parameters.

FinBERT is fine-tuned using a substantial corpus of 4.9 billion textual tokens extracted from corporate and analyst reports, as well as earnings call transcripts, notably leveraging the \textit{Financial Phrasebank} corpus \cite{malo2014good}. This fine-tuning process enables FinBERT to discern financial sentiment within textual inputs. The model classifies input text into one of three categories: positive, negative, or neutral, reflecting the underlying financial implications conveyed in the text. Positive sentiment typically aligns with positive market trends, while negative sentiment may indicate market downturns, and neutral sentiment suggests market indifference.

\begin{figure}[h!]
\centering
\includegraphics[width=\columnwidth]{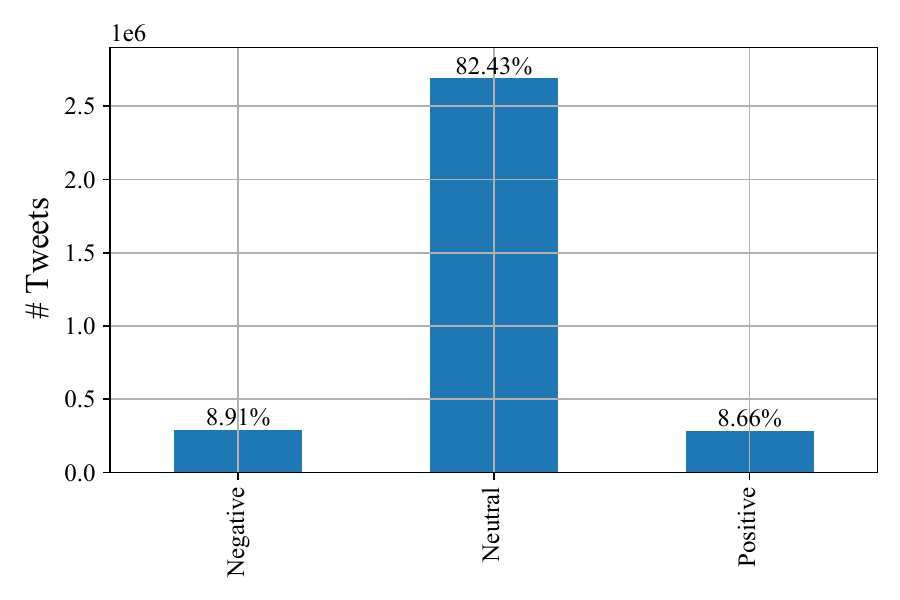}
\caption[]{FinBERT sentiment tweets classification.}
\label{tweetsClassification}
\end{figure}

In the context of analyzing StockTwits tweets related to the SPY ETF, depicted in Figure \ref{tweetsClassification}, FinBERT exhibits an imbalance in classification. Approximately $80\%$ of tweets are classified as neutral, while roughly $8\%$ are identified as carrying discernible financial sentiment, encompassing both positive and negative sentiments. This outcome aligns with expectations, given the nature of publicly available social media data, where genuine financial insights are relatively scarce. Nevertheless, even neutral classifications hold significance, as they reflect market sentiment and hype. Additionally, the volume of tweets on a particular day serves as valuable information in itself.

To quantify sentiment, akin to the approach proposed in \cite{hiew2019bert}, we compute a sentiment score $S_t$ using
\begin{equation}
\label{sentimentIndex}
    S_t = \frac{P_t -N_t}{T_t},
\end{equation}
where $P_t$ represents the number of positive tweets, $N_t$ signifies the number of negative tweets, and $T_t$ denotes the total number of tweets (including those labeled as neutral) on day $t$. This sentiment score provides a concise metric to gauge overall sentiment trends within the analyzed dataset.

\section{BAYESIAN-OPTIMIZED RECURSIVE FEATURE ELIMINATION}
\label{BORFESECT}
This section delineates the Bayesian-optimized Recursive Feature Elimination (BO-RFE) methodology. This automated algorithm enables the extraction of an optimal subset of features from a feature collection, ensuring maximal informativeness on the dataset while adhering to a predefined cost function.

Recursive Feature Elimination (RFE) operates with two configurable parameters: the target number of features to retain from the original set and the choice of kernel algorithm for feature selection. It is essential that the kernel algorithm produces a linear ranking of features; therefore, algorithms like SVM with radial basis function kernels are unsuitable for RFE due to their non-linear feature ranking.

The workflow of the algorithm is as follows: after selecting the kernel, it is trained, and feature importance is assessed based on the linear ranking. The least important features are iteratively removed, and the kernel algorithm is re-trained with the remaining features until the desired number of features is achieved.

Our approach merges RFE with Bayesian Optimization (BO). RFE enables the identification of the optimal number of features $\gamma$ and the set of dominant features $\Gamma$, given a target feature count and a kernel model. $\gamma$, a hyperparameter for RFE, is tuned alongside the parameter set $\Theta$ of the kernel model. BO automatically tunes $\gamma$ (and thereby $\Gamma$) and $\Theta$. The objective function for BO is the F1-score obtained on a reduced test set spanning $30$ days ($T^{test}{BO}$), using the same prediction model employed in the simulation as outlined in the literature \cite{liu2023stock}, after training on $222$ days ($T^{train}{BO}$). Bayesian Optimization is a widely recognized hyperparameter tuning approach; for detailed insights, please consult \cite{lim2}. 

\textit{Remark}: This procedure occurs within the training/validation set described in Section \ref{numericalSim}. The values of $\gamma_k$ and $\Theta_k$ will maximize, along the BO iterations $k=1,...,K$, the F1-score of the return sign prediction, eventually converging at the best $\gamma_K$, $\Gamma_K$ and $\Theta_K$.

\begin{algorithm}[H]
	\caption{Bayesian-optimized Recursive Feature Elimination Algorithm (BO-RFE)}
	\label{alg:MYALG}
	\begin{algorithmic}[1]	
		\For {$k = 1$,...,$K$}
        \State $\gamma_k, \Gamma_k = RFE(\Theta_k)$.
            \For {$\tau = 1$,...,$T^{test}_{BO}$}
            \State $\hat{y}_{\tau} = SVM(\gamma_k, \Gamma_k)$
            \EndFor
        \State F1-score$([y_1,...,y_{T^{test}_{BO}}],[\hat{y}_1,...,\hat{y}_{T^{test}_{BO}}])$
        \EndFor
        \State Return $\gamma_K, \Gamma_K$
	\end{algorithmic}
\end{algorithm}

Algorithm \ref{alg:MYALG} delineates the BO-RFE process. Here, $k$ denotes the BO iteration index, while $\tau$ represents the relative time index within the reduced test set. $y_{\tau}$ and $\hat{y}_{\tau}$ signify the true and predicted return signs at time $\tau$, respectively.

The RFE kernel is a \textit{random forest classifier}, with the optimally identified hyperparameter $\Theta$ representing the \textit{number of trees} within the forest. Random forest classifiers are established machine learning models wherein the predicted class is determined through a democratic election based on the predictions from $\Theta_K$ trees within the forest. Further insights into random forest classifiers can be found in \cite{parmar2019review}.

For BO convergence, we set $K$ to $50$, which proves to be sufficient. The initial feature set size is $10$, consistent with the feature set utilized in \cite{liu2023stock}, summarized in Table \ref{tab: features}.

The iterative procedure yields an optimal number of trees $\Theta_K = 5$, an optimal number of features $\gamma_K = 2$, and the corresponding optimal feature set $\Gamma_K = [R_{t-1}, S_t$ pre-market$]$. The achieved F1-score on the reduced test set $T^{test}_{BO}$ stands at $75\%$.

\section{CORRELATION ANALYSIS}
\label{correlation}
In this section we present the correlation analysis that allowed us to prove the meaningfulness of additional candidate regressors to the subset identified by the \textit{BO-RFE} algorithm. Table \ref{tab: pearsons} shows the Pearson correlation coefficient between the number of negative $N_t$, neutral $n_t$ and positive $P_t$ FinBERT-classified StockTwits tweets and SPY return $R_t$ and traded volume $V_t$, at the same day $t$. We highlight that negative tweets carry higher information content with respect to positive and neutral tweets. This reflects the interpretable behaviour that in negative scenarios the number of financially negative tweets increases, describing market hype-to-sell and ultimately leads to increased traded stock volumes and lower returns. This analysis motivates our choice of including in the regressors set the number of negative FinBERT-classified tweets.
\begin{table}[h!]
\centering
\renewcommand{\arraystretch}{1.1}
\begin{tabular}{ |c|c|c|c| } 
 \hline
 \textbf{Pearson [-]} & \# Tweets (-) $N_t$ &  \# Tweets (=) $n_t$ & \# Tweets (+) $P_t$\\
 \hline
 $R_t$  & $-0.06$ & $-0.04$ & $0.035$ \\
 $V_t$  & $0.64$ & $0.57$ & $0.55$ \\
 \hline
\end{tabular}
 \caption{Pearson correlation coefficient between number of negative, neutral and positive tweets (FinBERT labeled) and return and stock volume, at the same time instant.}
\label{tab: pearsons}
\end{table}

\begin{figure}[h!]
\centering
\includegraphics[width=1.05\columnwidth]{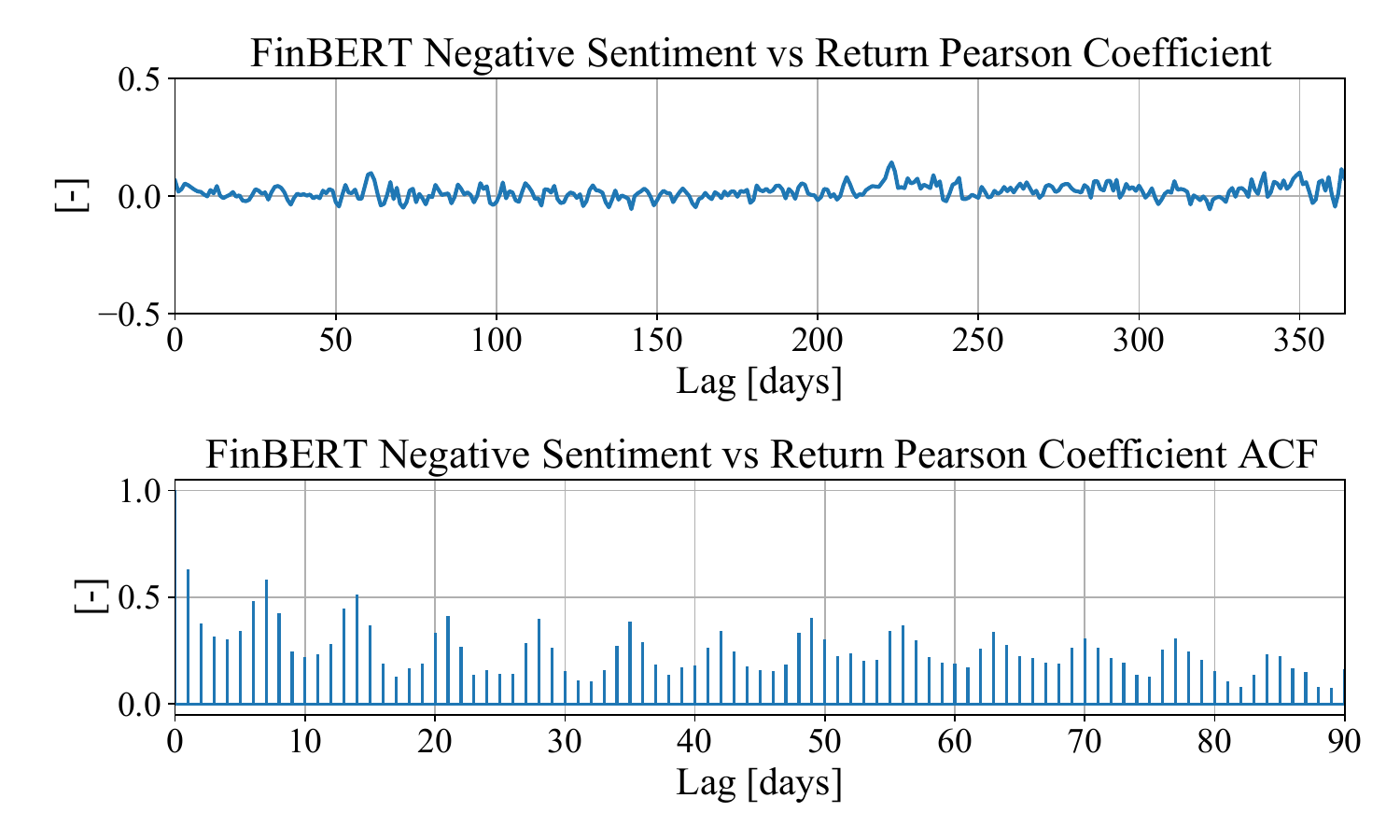}
\caption[]{Upper panel: FinBERT negative sentiment vs return Pearson correlation coefficient for different time lags. Lower panel: Pearson autocorrelation function.}
\label{pearsonLags}
\end{figure}
Figure \ref{pearsonLags} upper panel shows the Pearson correlation coefficient between FinBERT negative sentiment and SPY return for different time lags. In the lower panel, the Pearson autocorrelation function (ACF) is shown. Both upper and lower panel depicts a periodic behaviour in the Pearson coefficient for various lags. In particular the ACF reflects a weekly seasonality in the correlation between sentiment and returns with periodic peaks. This allows us to conclude that impact of sentiment information on return has a slow dynamics with weekly period. This analysis, jointly with the results in Table \ref{tab: pearsons} lead us to the choice of the additional set of regressors:
\begin{itemize}
    \item $S_{t-7}$ intra-market, to capture the seasonal behaviour of sentiment index in intra-market time.
    \item $S_{t-7}$ post-market, to capture the seasonal behaviour of sentiment index in post-market time.
    \item $N_{t-7}$, to capture the seasonal behaviour of number of negative tweets.
\end{itemize}
These 3 additional regressors, together with the ones identified through the BO-RFE procedure, constitute the set we propose as best configuration for the return prediction problem of SPY ETF.

For sake of compactness, we summarize in Table \ref{tab: features} the feature sets for the 3 strategies: 
\begin{table}[h!]
\centering
\renewcommand{\arraystretch}{1.1}
\begin{tabular}{ |l|c|c|c| } 
 \hline
 \textbf{Strategy} & Literature & BO-RFE - 2 & BO-RFE - 5 \\
 \hline
 $R_{t-1}$  & \textbf{yes} & \textbf{yes} & \textbf{yes} \\
 $S_t$ pre-market & \textbf{yes} & \textbf{yes} & \textbf{yes}\\
 $S_{t-1}$ intra-market & \textbf{yes} & no & no\\
 $S_{t-7}$ intra-market & no & no & \textbf{yes}\\
 $S_{t-1}$ post-market & \textbf{yes} & no & no\\
 $S_{t-7}$ post-market & no & no & \textbf{yes}\\
 $F_{t-1}$ [$6$ financial features]& \textbf{yes} & no & no\\
 $N_{t-7}$ & no & no & \textbf{yes}\\
 \hline
\end{tabular}
 \caption{Feature sets for \textit{Literature}, \textit{BO-RFE-2} and \textit{BO-RFE-5} algorithms.}
\label{tab: features}
\end{table}
$F_{t-1}$ represents a synthetic notation for the $6$ financial features used exclusively in \textit{Literature} startegy, taken at day $t-1$ with the goal of predicting the return sign at time $t$. The financial features used in literature are: opening, closing, high and low prices, stock volume and trade value. For additional details see \cite{liu2023stock}.

\section{THE PREDICTION MODEL}
\label{prediction}
The prediction model is the same used in \cite{liu2023stock} and it is used both in the BO-RFE algorithm \ref{alg:MYALG} and during online return sign predictions. It consists in a support vector machine (SVM) with the combination of synthetic minority over-sampling technique (SMOTE, see \cite{fernandez2018smote}) to address class imbalance (negative return sign days are around $300$ whereas positive days are around $420$) and bagging (a procedure that performs bootstrapping on original data and trains $L$ classifiers in parallel to eventually generate a democratic ensemble classifier for return sign prediction, resembling the random forest classifier concept, see \cite{skurichina1998bagging}).

The SVM (SVC if we consider classification) training problem is the following.
\begin{subequations}
	\label{SVC}
	\begin{equation}
            \label{cost}
		\operatorname{min}_{w,b,\xi}\{\frac{1}{2}w^T \cdot w + C \sum^n_{i=1}\xi_i\}
	\end{equation}    
	\begin{equation}
		y_i(w^T\cdot \phi(x_i)+b) \geq 1 - \xi_i 
	\end{equation}
        \begin{equation}
		\xi_i \geq 0 \,\,\, \forall i=1,...,n 
	\end{equation}
        \begin{equation}
		x_i,w \in R^P \,\,\, \forall i=1,...,n 
	\end{equation}
        \begin{equation}
		  b \in R, \xi_i \in R^n \,\,\, \forall i=1,...,n 
	\end{equation}
\end{subequations}
Where $P$ is the number of features, $n$ is the number of samples, $x_i$ is the i-th input of $P$ features among the $n$ samples, $w$ are the model weights, $b$ is the bias, $\xi$ is a slack variable for weakly constraining the hyperplanes, $C$ and $\phi(x_i)$ (kernel) are SVC hyperparameters. Following \cite{cortes1995support}, the machine implements the idea that input vectors are non-linearly mapped to a very high-dimension feature space. In this space a separating hyperplane is constructed, allowing to distinguish samples among different target classes exploiting non-linear relationships which otherwise would have been hidden. The model degrees of freedom depend on $C$, the penalty weight of the slack variables, and $\phi(x_i)$, the kernel of the SVM. In this application, we strictly stick with the hyperparameter choices $C=1$ and \textit{radial basis function} kernel, carried on in \cite{liu2023stock}, to provide a fair comparison for the same SVM model. We highlight that the number of model parameters (in particular model weights $w$) depends on the number of features $P$. For this reason, a lower number of features as in the case of \textit{BO-RFE-2} and \textit{BO-RFE-5} ultimately leads to a lower amount of model parameters, thus requiring a smaller training dataset.

\section{EXPERIMENTAL RESULTS}
\label{numericalSim}
We perform the numerical simulations with real StockTwits and SPY ETF data from 2020-03-24 to 2022-02-13. With the purpose of aligning StockTwits tweet data (available each day) and the financial return data (available only during work days) we carried on a linear interpolation during weekend days on financial data. This allows us to capture the textual data available in the weekends that can effectively affect the market during the work days, due to financial news made available during saturday and sunday. However, during the online trading simulation the trading actions (buy/sell) will be applied exclusively during work days. All computations are carried out on an Intel Core i7-8750H with 6 cores, at 2.20 GHz (maximum single core frequency: 4.10 GHz), with 16 GB RAM. Code is implemented in Python 3.7.11. Training and validation sets cover the $84\%$ of the data. Test set is the remaining $16\%$. Training and validation sets are merged since training is done in a moving window (with size $W$) fashion. Table \ref{tab: numerical settings} collects each strategy setting. The only hyperparameter tuned for each strategy is the moving window size $W$. For each strategy, we carried on a sensitivity analysis on the training/validation sets for different window sizes, picking up the best window $W$ for each strategy. \textit{Literature} best window size confirms the range described in \cite{liu2023stock} where the authors tested the window range between $232$ and $252$. Indeed, the identified best window range for \textit{Literature} is $240$. For \textit{BO-RFE-2} and \textit{BO-RFE-5} strategies instead, the best window size is $40$ and $210$, respectively. This window size reduction depends on the lower number of regressors involved in BO-RFE strategies. Following SVM model structure, a lower number of features ultimately implies a lower number of parameters and hence a lower optimal number of data required in training. This allows us to conclude that \textit{BO-RFE-5} strategy requires 1 month less of training data with respect to its literature benchmark. This is an interesting point for an online application of the algorithm as it requires less training data (and therefore a shorter warm-up period), moreover, it guarantees higher reactivity properties given the shorter required past horizon.
\begin{table}[h!]
\centering
\renewcommand{\arraystretch}{1.1}
\begin{tabular}{ |c|c|c|c| } 
 \hline
 \textbf{Strategy} & Literature & BO-RFE - 2 & BO-RFE - 5 \\
 \hline
 Window $W$ [days]  & $240$ & $40$ & $210$ \\
 \# Features & $10$ & $2$ & $5$  \\
 \# Financial Features & $7$ & $1$ & $1$ \\
 \# Sentiment Features & $3$ & $1$ & $4$ \\
 \hline
\end{tabular}
 \caption{Strategy settings - Hyperparameters and number of (financial and sentiment) features.}
\label{tab: numerical settings}
\end{table}

We remark that the contribution of sentiment-based features is relevant in \textit{BO-RFE} strategies. Indeed, they constitute the $50 \%$ of the features in \textit{BO-RFE-2} and the $80 \%$ in \textit{BO-RFE-5}, differently from \textit{Literature} where the sentiment features are only the $30 \%$ of the total. Given the promising results of \textit{BO-RFE}, this proves the huge impact of sentiment-based information, processed through FinBERT transformer architecture, to generate meaningful regressors for financial return series prediction.

\subsection{RETURN SIGN PREDICTION}
\label{accuracy}
\begin{figure}[h!]
\centering
\includegraphics[width=\columnwidth]{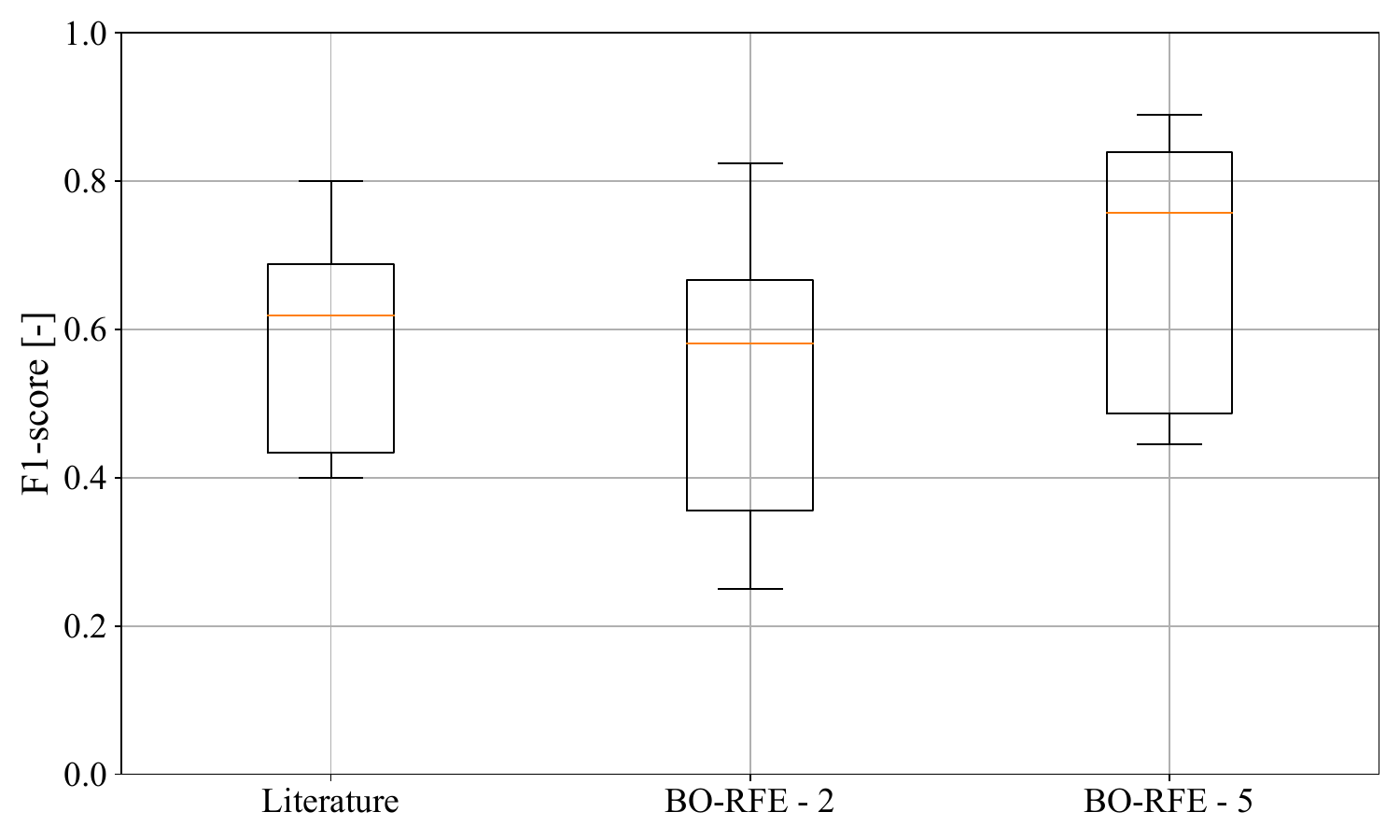}
\caption[]{F1-score on $8$ batches of $10$ trading days each in test set.}
\label{boxplot_profitto}
\end{figure}

Table \ref{tab: metrics} collects accuracy, precision, recall and F1-score for the 3 strategies in the test set of $80$ trading days ($110$ days, including non-trading weekends). \textit{BO-RFE-5} achieves outstanding performances with an accuracy (e.g., percentage of correctly labeled data, $accuracy = \frac{TP+TN}{TP+TN+FP+FN}$) of the $64.1\%$, a precision (e.g., percentage of correctly labeled data over the total labeled, for each class, $precision = \frac{TP}{TP+FP}$) above $63\%$, a recall (e.g., percentage of correctly labeled data over the total number of data belonging to the same class, $recall = \frac{TP}{TP+FN}$) above $78\%$ and a F1-score (e.g., synthetic measure of precision and recall, F1-score $ = \frac{2 \cdot precision \cdot recall}{precision + recall}$) above $70\%$.
\begin{table}[h!]
\centering
\renewcommand{\arraystretch}{1.1}
\begin{tabular}{ |c|c|c|c| } 
 \hline
\textbf{Strategy} & Literature & BO-RFE - 2 & BO-RFE - 5 \\
 \hline
 Accuracy & $0.564$ & $0.512$ & $0.641$ \\
 Precision & $0.583$ & $0.541$ & $0.634$  \\
 Recall & $0.667$ & $0.619$ & $0.785$ \\
 F1-score  & $0.622$ & $0.577$ & $0.702$ \\
 \hline
\end{tabular}
 \caption{Classification metrics in test set for Literature, BO-RFE - 2 and BO-RFE - 5 architectures.}
\label{tab: metrics}
\end{table}
Figure \ref{boxplot_profitto} shows the test set F1-score, for $8$ batches of $10$ trading days ($2$ weeks including weekends). \textit{BO-RFE-5} outstands \textit{Literature} and \textit{BO-RFE-2}. Its scores show higher mean and quartiles, demonstrating a consistent improvement over all the test set. \textit{Literature} scores confirm those described in \cite{liu2023stock}.
These results demonstrate that the careful choice of features, considering the most significant ones identified in the literature and a correlation analysis that evaluates the potential contribution of past regressors, is decisive in prediction quality. 
The results in particular prove that the contribution of the regressors $S_{t-7}$ intra-market, $S_{t-7}$ post-market, and \textit{$N_{t-7}$} is significant. Compared to the BO-RFE-2 feature set, BO-RFE-5 regressors significantly improve performance, demonstrating the information content of the sentiment information at lag $7$ and the slow dynamics of user social media interactions on the financial asset. Furthermore, it confirms the informativeness of using not only the sentiment index $S_t$ described in equation \eqref{sentimentIndex} but also the number of negative FinBERT-classified tweets in predicting returns.
\subsection{TRADING SIMULATION}
\label{simulation}
In this section we apply the strategies in a trading simulation. We show that, for the same return sign prediction model, the profit magnitude depends on the quality of the feature selection. The trading strategy is the same described in \cite{cestari2023hawkes} and it is as follows: if the predicted sign is positive (negative), the automatic trading algorithm will purchase (short-sell) SPY ETF for a value of $10.000$ \$. If the actual sign matches the predicted one, the trader has a profit; otherwise, the trader incurs in a loss. No transaction costs are considered.
\begin{figure}[h!]
\centering
\includegraphics[width=\columnwidth]{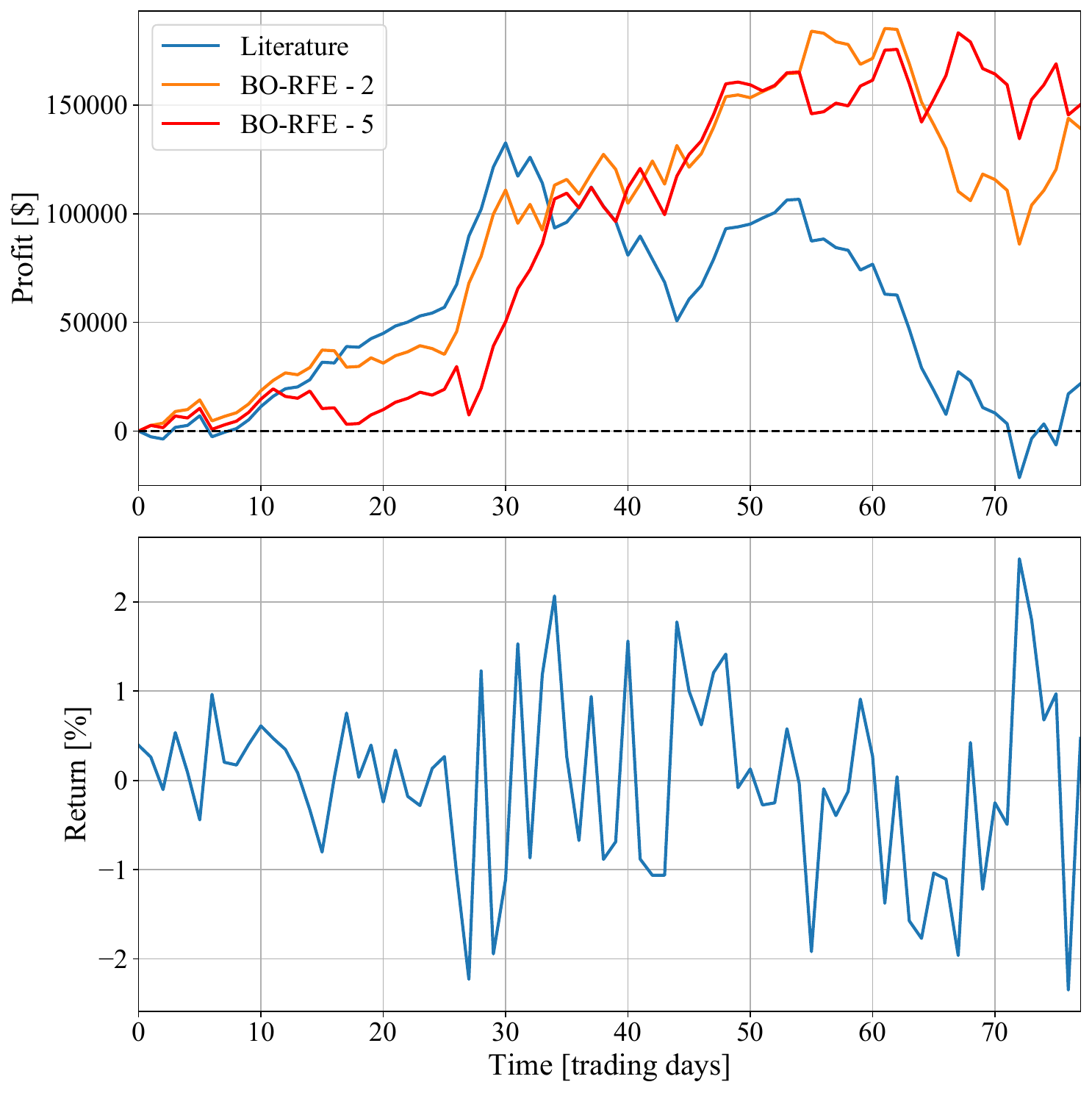}
\caption[]{Upper panel: cumulative profit time series in test set. Lower panel: SPY return.}
\label{timeseriesProfit}
\end{figure}
Figure \ref{timeseriesProfit}, upper panel, shows the cumulative profit time series obtained with the different trading algorithms in the test set. Before day $30$, return absolute value is bounded (as it is clear in the lower panel) below $1\%$. The strategies have lower profit and \textit{Literature} outperforms. From day $30$ onward, the return absoulte value is higher (with higher oscillations around the mean, e.g., higher volatility). This allows the strategies to cumulate profit, proportionally to their accuracy in predicting the return sign. This ultimately leads to outstanding performance for \textit{BO-RFE-5}. The ranking shows that both the strategies \textit{BO-RFE-2} and \textit{BO-RFE-5} have higher profit. This depends on their higher accuracy earned in the more volatile period, whereas the \textit{Literature} model accuracy is "wasted" in the beginning of the simulation horizon, where the returns are lower and, eventually leads to a lower cumulative profit, slightly above $10.000$ \$ in the test set considered.

\section{CONCLUSIONS}
\label{conclusions}
In this study, we explore the influence of feature selection on the performance of return sign prediction for the SPY, an ETF tied to the $S\&P$ 500 index, using the same prediction model. Employing an automated feature selection technique, BO-RFE, we combine the Recursive Feature Elimination (RFE) procedure with Bayesian Optimization (BO) to extract an optimal subset of features from those identified in existing literature.

We conduct a correlation analysis to unveil the sluggish, weekly seasonality-driven dynamics of sentiment effects on returns, particularly focusing on sentiments labeled as negative by the LLM FinBERT. Integrating this insight as additional regressors, we demonstrate how the choice of input features significantly impacts prediction quality and, consequently, achievable profits with different trading strategies on a test set. Notably, the \textit{BO-RFE-5} solution exhibits a substantial performance enhancement.

Our findings underscore the predominance of sentiment-based features among the optimal set, \textit{highlighting the significant influence of textual information on financial assets}. While the state-of-the-art solution comprises 3 out of 10 sentiment-based features, our \textit{BO-RFE-5} method increases this ratio to 4 out of 5. Additionally, \textit{the selection of "essential" features leads to model simplification, resulting in a smaller optimal training window and, consequently, enhanced responsiveness and shorter warm-up periods}.

Future endeavors will involve validating our proposed methodology across a broader testing horizon and exploring its applicability to different financial assets.

\bibliographystyle{IEEEtran}
\bibliography{bibliography}

\addtolength{\textheight}{-12cm}   


\end{document}